\documentclass[reprint, superscriptaddress, secnumarabic, amssymb, nobibnotes, aps, prl]{revtex4-2}

\pdfoutput=1

\usepackage{graphicx}
\usepackage{epstopdf}
\usepackage[T1]{fontenc}
\usepackage[utf8]{inputenc}
\DeclareUnicodeCharacter{2212}{~}
\usepackage{amsbsy}
\usepackage{gensymb}
\usepackage[T1]{fontenc}
\usepackage{amsmath}
\usepackage{amssymb}
\usepackage{bbm}
\usepackage{physics}
\usepackage{braket}
\usepackage{xcolor}
\allowdisplaybreaks
\usepackage{graphicx}
\usepackage[colorlinks=true]{hyperref}  
\hypersetup{
    bookmarks=true,         % show bookmarks bar?
    unicode=false,          % non-Latin characters 
    pdftoolbar=true,        % show Acrobat
    pdfmenubar=true,        % show Acrobat 
    pdffitwindow=false,     % window fit to page when opened
    pdfstartview={FitH},    % fits the width of the page to the window
    pdftitle={Fully gapped superconductivity in centrosymmetric and non-centrosymmetric  Re-B compounds probed by $\mu$SR},    % title
    pdfauthor={S. Sharma, Arushi, J. Beare, M. Nugent, M. Pula, T. Munsie, R. P. Singh, G.M. luke}, %aut
    pdfsubject={},   % subject of the document
    pdfcreator={},   % creator of the document
    pdfproducer={}, % producer of the document
    pdfkeywords={} {} {}, % list of keywords
    pdfnewwindow=true,      % links in new window
    colorlinks=true,       % false: boxed links; true: colored links
    linkcolor=blue, %red,          % color of internal links (change box color with linkbordercolor)
    citecolor=blue,        % color of links to bibliography
    filecolor=magenta,      % color of file links
    urlcolor=blue           % color of external links
} 
\usepackage[normalem]{ulem}

\renewcommand{\approx}{\simeq}

\begin{document}
    
\preprint{APS/123-QED}

\title{Fully gapped superconductivity in centrosymmetric and non-centrosymmetric  Re-B compounds probed with $\mu$SR}% Force line breaks with \\

\author{S.~ Sharma}
\affiliation{Department of Physics and Astronomy, McMaster University, Hamilton, Ontario L8S 4M1, Canada}
\author{Arushi}
\affiliation{Department of Physics, Indian Institute of Science Education and Research Bhopal, Bhopal, 462066, India}
\author{K. Motla}
\affiliation{Department of Physics, Indian Institute of Science Education and Research Bhopal, Bhopal, 462066, India}
\author{J.~Beare}
\affiliation{Department of Physics and Astronomy, McMaster University, Hamilton, Ontario L8S 4M1, Canada}
\author{M.~Nugent}
\affiliation{Department of Physics and Astronomy, McMaster University, Hamilton, Ontario L8S 4M1, Canada}
\author{M.~Pula}
\affiliation{Department of Physics and Astronomy, McMaster University, Hamilton, Ontario L8S 4M1, Canada}
\author{T.J.~Munsie}
\affiliation{Defence Research and Development Canada, Suffield Research Centre, Ralston, Alberta  T0J 2N0, Canada
}
\author{A. D. Hillier}
\affiliation{ISIS Facility, STFC Rutherford Appleton Laboratory,
Harwell Science and Innovation Campus, Oxfordshire, OX11 0QX, UK}
\author{R.~P.~Singh}
\affiliation{Department of Physics, Indian Institute of Science Education and Research Bhopal, Bhopal, 462066, India}
\author{G.~M.~Luke}
\email[]{luke@mcmaster.ca}
\affiliation{Department of Physics and Astronomy, McMaster University, Hamilton, Ontario L8S 4M1, Canada}
\affiliation{TRIUMF, Vancouver, British Columbia V6T 2A3, Canada}

\date{\today}% It is always \today, today,

\begin{abstract}
\begin{flushleft}
\end{flushleft}

We present a comprehensive study on superconducting properties of Re$_7$B$_3$ and Re$_3$B through specific heat, magnetic susceptibility, resistivity, and transverse and zero-field muon spin rotation/relaxation ($\mu$SR) experiments on polycrystalline samples. 
  Re$_7$B$_3$ (T$_C$ = 3.2~K) is a non-centrosymmetric type-II ($\kappa$ $\approx$ 9.27) superconductor in the weak coupling ($\lambda_{e-ph}$ = 0.54) regime.  On the other hand, Re$_3$B (T$_C$ = 5.19~K) is a centrosymmetric type-II ($\kappa$ $\approx$ 34.55) superconductor in the moderate coupling ($\lambda_{e-ph}$ = 0.64) regime. Our transverse-field $\mu$SR measurements show evidence for isotropically gapped BCS type superconductivity with normalized gap ($\Delta_0/k_BT_C$) values of 1.69 (Re$_7$B$_3$) and 1.75 (Re$_3$B).                                                                                                                                                                                                                                                                                                                                                  
\end{abstract}
\maketitle
	
%\tableofcontents
%%%%%%%%%%%%%%%%%%%%%%%%%%%%%%%%%%%%%%%%%%%%%%%%%%%%%%%%%%%%%%%%%%%%%%%%%%%%%%%%%%%%%%%%
\section{\label{sec:level1}INTRODUCTION \protect\\ }
Symmetry breaking in phase transitions is a central concept in physics and superconductivity is a good phenomenon which  demonstrates it. Conventional superconductors break gauge symmetry while unconventional superfluids and superconductors can break other kinds of symmetries.  It was after the discovery of unconventional superconductivity in inversion symmetry lacking CePt$_3$Si \cite{Bauer2004}, that non-centrosymmetric superconductors (NCS) jumped into the limelight. The lack of inversion symmetry gives rise to an anti-symmetric spin-orbital coupling, which can lead to an admixture of the spin-singlet and spin-triplet Cooper pairs \cite{Smidman2017,Gorkov2001,Yuan2006}. The mixed-pairing state can prompt non-centrosymmetric superconductors to manifest remarkably different properties from conventional BCS superconductors: for example, point or line nodes in the superconducting gap function \cite{Takeya2007,Chen2011,Kuroiwa2007,Shao2018}, presence of multiple superconducting gaps \cite{Kuroiwa2007,Shang2019}, Pauli limit exceeding upper critical fields \cite{Karki2010,Bauer2004,Bao2015,Kimura2005,Arushi2020}, topologically protected zero-energy surface bands \cite{Brydon2011}, and  time-reversal symmetry (TRS) breaking \cite{Hillier2009,Singh2018}. The non-centrosymmetricity  and existence of TRS breaking in  La$_7$Ir$_3$ \cite{TBarker2015} makes a good case  for iso-structural Re$_7$B$_3$ to show TRS breaking and hence, unconventional superconductivity. 

There have been many cases of Re based non-centrosymmetric superconductors exhibiting unconventional superconducting properties \cite{Singh2013, Singh2017, Singh2018} . The recent discovery of TRS breaking fields in centrosymmetric elemental Re \cite{ShangRe} has ignited fresh interest in Re based superconductors. Since the TRS breaking in ReT (T = transition metal) alloys is attributed to Re, Re$_3$B (centrosymmetric space group Cmcm) due to its high Re content, also becomes an interesting candidate for exhibiting TRS breaking. This can also help identify the existence of a critical Re concentration beyond which a compound will show TRS breaking. Through the study of these compounds, we can better understand the relative importance of Re concentration and non-centrosymmetricity in TRS breaking/unconventionality. In addition, this will help in creating a better understanding of the complex superconductivity \cite{Khan2016} in Re based superconductors. 
    
Although there have been some studies using micro and macroscopic techniques in their normal state \cite{Lue2008} as well as superconducting state \cite{Takagiwa2003,Kawano2003,Strukova2001}, there is not much information about the symmetry of the superconducting gap and possible presence of TRS breaking fields of Re$_7$B$_3$ and Re$_3$B. In this paper, we report the superconducting properties of these compounds, through magnetization, resistivity, specific heat, and muon spin rotation/relaxation ($\mu$SR) measurements. 
%%%%%%%%%%%%%%%%%%%%%%%%%%%%%%%%%%%%%%%%%%%%%%%%%%%%%
\section{EXPERIMENTAL DETAILS}
Our samples were synthesized by repeatedly arc melting Re (99.9\%) and 5\% excess B (99.9\%) in a high purity argon atmosphere carefully flipping the ingot before each melting. A small portion of the sample was powdered and characterized with an X'pert PANalytical x-ray diffractometer (Cu-K$\alpha$ radiation, $\lambda$ = 1.540~\AA) to confirm the phase purity of the samples.

The superconducting properties of both Re$_7$B$_3$ and Re$_3$B were measured using specific heat, magnetization, resistivity, and muon spin relaxation/rotation ($\mu$SR) measurements.   Magnetization measurements were carried out in Quantum Design MPMS-3 SQUID magnetometer \cite{QuantumDesignUSA2016}. The specific heat and resistivity data were obtained using a Quantum Design Physical Property Measurement System (PPMS).  
$\mu$SR measurements were performed at the M15 beamline of TRIUMF's Center for Molecular and Materials Science, Vancouver, which was equipped with a dilution refrigerator. In the transverse-field geometry, the spins of the implanted muons are initially perpendicular to the magnetic field applied to the sample. At the M15 beamline, the magnetic field was parallel to the beam axis, and therefore, to achieve TF geometry, the muon spins were rotated before implantation, leading to the use of left and right detectors for asymmetry measurement as a function of time. The polycrystalline samples were cut into plates with a diamond saw before mounting on the silver sample holder (cold finger). Copper grease was used to fix the plates on the sample holder which was wrapped by a thin silver foil to ensure good thermal contact.
\begin{figure}[]
\includegraphics[width=91.5mm]{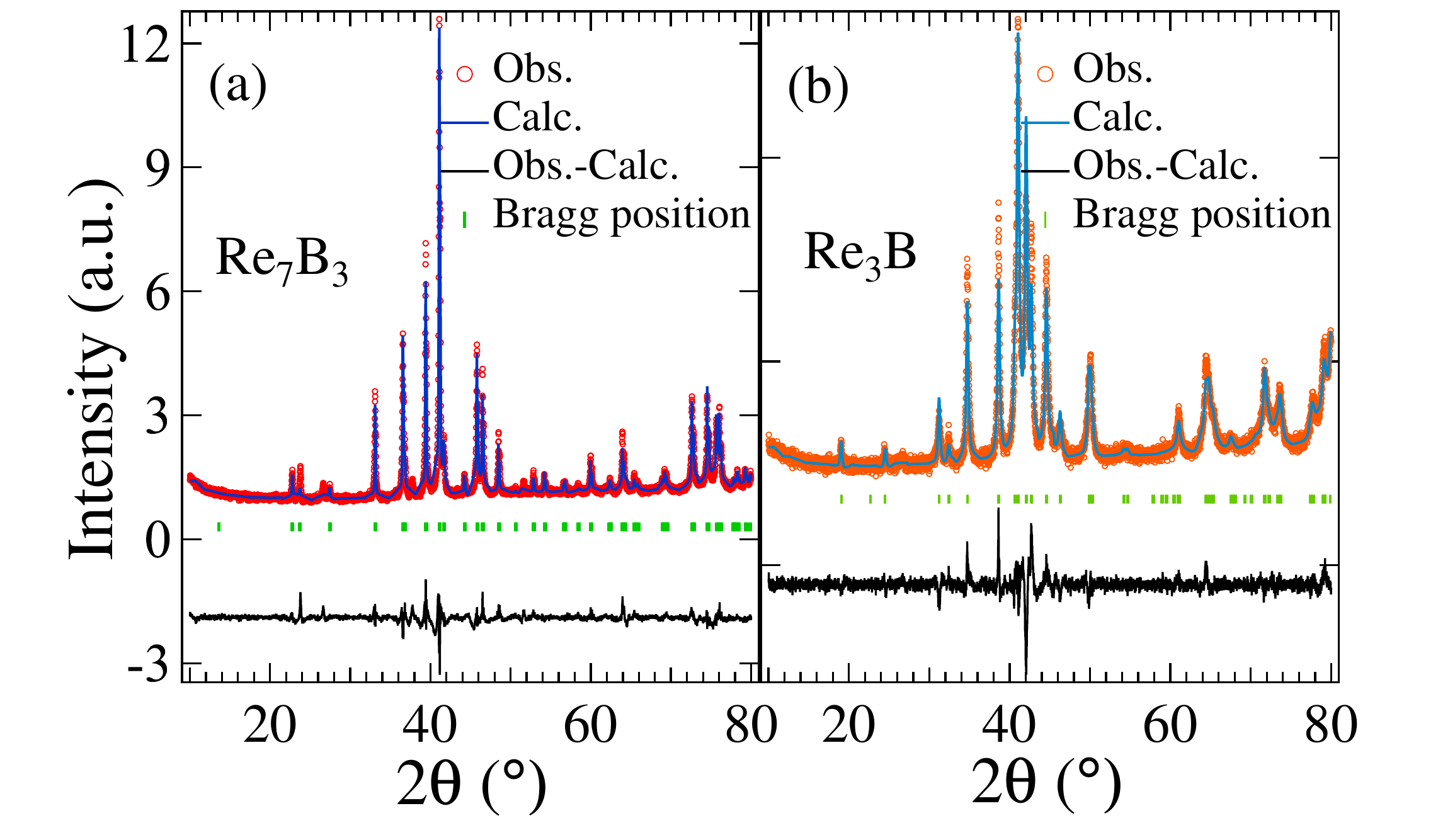}% Here is how to import EPS art
\caption{\label{xrd}  
The observed and calculated intensities from powder XRD experiment for (a) Re$_3$B and (b) Re$_7$B$_3$ along with Bragg positions are plotted here. 
}
\end{figure}

We performed zero-field muon spin relaxation experiments to detect possible spontaneous fields in the superconducting state. This required zeroing of the stray field into the sample space, which involved very accurate measurement of the field and subsequent application of the negative field by perpendicular magnets. The zero-field environment (accurate to 2~$\mu$T) was achieved by following the procedure described by Morris et. al. \cite{Morris2003}.

%%%%%%%%%%%%%%%%%%%%%%%%%%%%%%%%%%%%%%%%%%%%%%%%%%%%%%%%%%%%%%%%%%%%%%%%%%%%%%%%%%%%%%%%%%%
\section{RESULTS AND DISCUSSION}
%%%%%%%%%%%%%%%%%%%%%%%%%%%%%%%%%%%%%%%%%%%%%%%%%%%%%%%%%%%%%%%%%%%%%%%%%%%%%%%%%%%%%%%%%%%%
\paragraph*{X-ray Diffraction(XRD)}
Fig \ref{xrd} shows the observed and calculated powder XRD patterns along with the difference and Bragg positions for the two superconductors. Our measurements confirm that Re$_3$B adopts the centrosymmetric space-group Cmcm (63)  while Re$_7$B$_3$ forms in non-centrosymmtric space-group P63mc (186). The good agreement of observed and calculated intensity is an indication of the phase purity of our samples. However, in Re$_7$B$_3$ , we see a roughly 3\% ReB$_2$ impurity phase present while an additional peak (corresponding to less than 3\% concentration) could not be indexed to any known Re/B phases.  No evidence for superconducting contributions from any of the impurity phases was seen in any of our measurements, and as such, these small impurity phases do not affect our measured superconducting properties.

\begin{figure}[t]
\includegraphics[width=91.5mm]{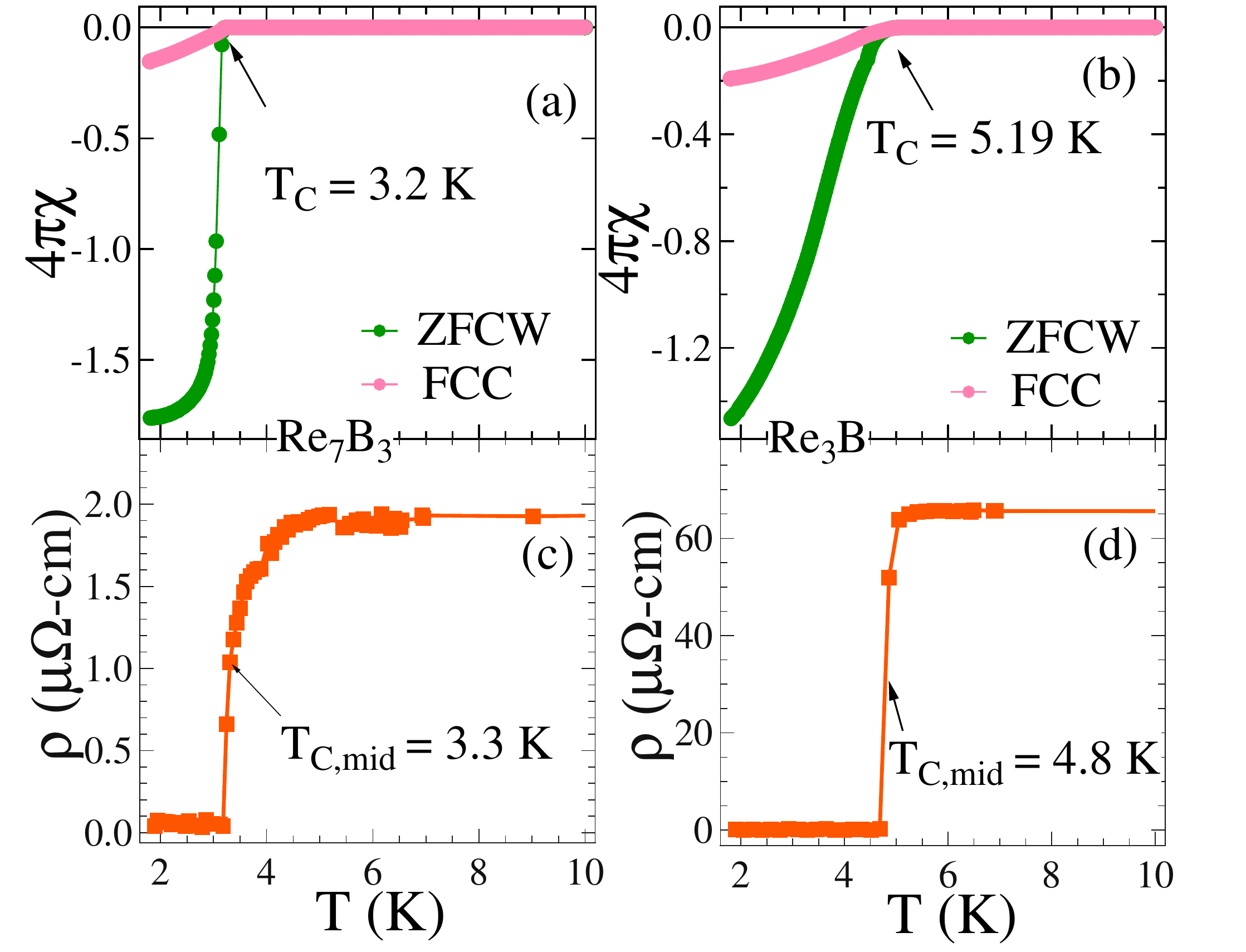}% Here is how to import EPS art
\caption{\label{Tc} The top half of the figure shows zero-field cooled warming (ZFCW) and field cooled cooling (FCC) volume magnetic susceptibility data in 10 G field for (a) Re$_7$B$_3$  and (b) Re$_3$B. The lower half of the figure shows zero-field resistivity data with superconducting transitions at 3.3~K and 4.8~K for (c) Re$_7$B$_3$ and (d)  Re$_3$B, respectively, slightly different from the values obtained through the magnetic susceptibility data.}
\end{figure}

\paragraph*{Resistivity}
Fig. \ref{Tc} (c-d) depicts the superconducting transitions on the resistivity data. T$_{C,mid}$ is the temperature at the midpoint of the resistivity drop. Re$_7$B$_3$ transitions to the superconducting state at T$_{C,mid}$ = 3.3~K while R$_3$B transitions at T$_{C,mid}$ = 4.8~K. We fit the normal state resistivity data with $\rho = \rho_0 + \rho_1T^2$ to compute the normal state residual resistivity, $\rho_0$, obtaining 1.86 $\pm$ 0.01~$\mu\Omega cm$ and 65.38 $\pm$ 0.09~$\mu\Omega cm$  for Re$_7$B$_3$ and Re$_3$B, respectively, which in turn was used to estimate the mean free path of the electrons, $l$. 

\paragraph*{Magnetization}

Fig. \ref{Tc} (a-b) depicts the magnetic susceptibility data with $T_{C}$ onset of 3.2~K and 5.19~K for Re$_7$B$_3$ and R$_3$B, respectively. The shielding volume fraction is close to 100 \% indicating the full spread of superconductivity within the samples.  The higher susceptibility in the ZFCW case implies the presence of flux pinning and the value lower than -1 likely reflects the lower sample density than the one used in the calculation. The Re$_3$B shows a broader transition in comparison to Re$_7$B$_3$. Since the same effect is mimicked by the specific heat measurements, the sample inhomogeneity or the presence of small disorder could be the reason for it.
%%%%%%%%%%%%%%%%%%%%%%%%%%%%%%%%%%%%%%%%%%%%%%%%%%%%%%%%%%%%%%%%%%%%%%%%%%%%%%%%%%%%%%%%%%%%%%%%
%%%%%%%%%%%%%%%%%%%%%%%%%%%%%%%%%%%%%%%%%%%%%%%%%%%%%%%%%%%%%%%%%%%%%%%%%%%%%%%%%%%%%%%%%%%%%%%%

\begin{figure}[t]
\includegraphics[width=91.5mm]{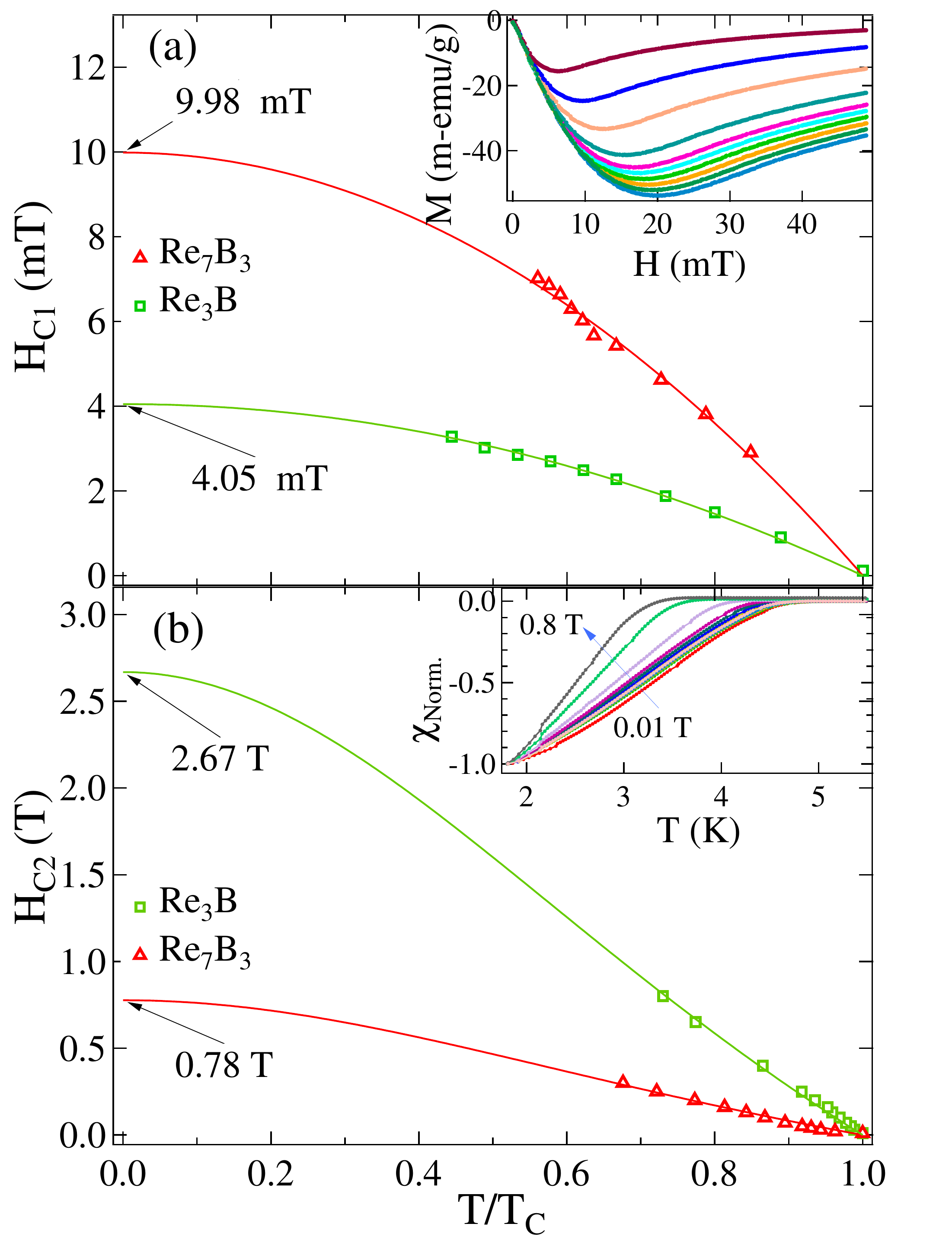}% Here is how to import EPS art
\caption{\label{Hc2}  
 (a). The lower critical field $H_{C1}$ values were determined from the magnetization (M) versus field (H) curves at various temperatures [Inset]. The red and green lines represent the $H_{C1}(T)$ data as modeled using the Ginzburg-Landau theory (Eq. \ref{Hc2Eq}). $H_{C1}$(0) values obtained from the fits are shown in the figure [Tagged arrow].
(b). Magnetic susceptibility ($\chi$) versus temperature (T) is plotted at various fields for Re$_3$B [Inset]. The upper critical field ($H_{C2}$) values as a function of temperature (T) as determined from $\chi$ versus T graph are plotted. The red and green line represent the fits of the data in accordance with the Ginzburg-Landau theory (Eq. ~\ref{Hc1Eq}).}
\end{figure}

The inset of the Fig. \ref{Hc2} shows the magnetization versus field curves at different temperatures. No demagnetization corrections were applied due to the somewhat irregular shape of the samples. The magnetic flux starts penetrating the sample at lower critical field (H$_{C1}$) which leads to non linear behavior of magnetization for fields greater than H$_{C1}$. The H$_{C1}$ was extracted from the magnetization isotherms and modeled using the Ginzburg-Landau Eq. \ref{Hc1Eq} to determine the H$_{C1}$(0), the lower critical field at 0 K. The H$_{C1}$(0) for Re$_7$B$_3$ and Re$_3$B were obtained to be 9.98 $\pm$ 0.08~mT and 4.05 $\pm$ 0.03~mT, respectively. There is significantly larger systematic uncertainty in these values due to uncertainty in the demagnetization factor. %for this reason estimate based on our muon spin rotation measurements (reported later in the paper) provide a more accurate estimate of H$_{C1}$.
\begin{equation}
\label{Hc1Eq}
H_{C1}(T) = H_{C1}(0)\left[ 1 - \left(\frac{T}{T_C}\right)^2\right]
\end{equation}
%%%%%%%%%%%%%%%%%%%%%%%%%%%%%%%%%%%%%%%%%%%%%%%%%%%%%%%%%%%%%%%%%%%%%%%%%%%%%%%%%%%%%%%%%%%%%%%%%
%%%%%%%%%%%%%%%%%%%%%%%%%%%%%%%%%%%%%%%%%%%%%%%%%%%%%%%%%%%%%%%%%%%%%%%%%%%%%%%%%%%%%%%%%%%%%%%%%
The magnetic susceptibility as a function of temperature for various applied fields for Re$_3$B is shown in the inset of Fig. \ref{Hc2} (a). Fig. \ref{Hc2} (b) shows the upper critical field values varying with reduced temperature which was extracted from the susceptibility curves for the respective samples. We used the Ginzburg-Landau model (Eq. \ref{Hc2Eq}), which is effective for temperatures close to T$_C$, to estimate the H$_{C2}$(0) as shown in Fig. \ref{Hc2} (b). The H$_{C2}$(0) for Re$_7$B$_3$ and Re$_3$B obtained along with their statistical errors are 0.78 $\pm$ 0.02~T and 2.67 $\pm$ 0.04~T, respectively.
\begin{equation}
\label{Hc2Eq}
H_{C2}(T) = H_{C2}(0)\left[ \frac{1 - t^2}{1 + t^2}\right],
\end{equation}
where t = T/T$_C$. H$_{C2}$ is related to the Ginzburg-Landau coherence length ($\xi_{GL}$) through  \cite{Tinkham1996},   
\begin{equation}
\label{CL}
H_{C2}(0) = \frac{\phi_0}{2\pi\xi_{GL}^2},
\end{equation}
where $\phi_0$ (=$2.07\times10^{-15}$ $Tm^{2}$) is the magnetic flux quantum. We calculated $\xi_{GL}$(0) using our $H_{C2}(0)$ values, obtaining 20.68 $\pm$ 0.12~nm   for  Re$_7$B$_3$ and 11.13 $\pm$ 0.08~nm  for Re$_3$B.  
The Ginzburg-Landau penetration depth ($\lambda_{GL}$) can also be calculated using the expression \cite{Tinkham1996},
\begin{equation}
    \label{PD}
    H_{C1}(0) = \frac{\phi_0}{4\pi\lambda^2_{GL}(0)}\left(\ln\frac{\lambda_{GL}(0)}{\xi_{GL}(0)}\right).
\end{equation}
%%%%%%%%%%%%%%%%%%%%%%%%%%%%%%%%%%%%%%%%%%%%%%%%%%%%%%%%%%%%%%%%%%%%%%%%%%%%%%%%%%%%%%%%%%%%%%%
%%%%%%%%%%%%%%%%%%%%%%%%%%%%%%%%%%%%%%%%%%%%%%%%%%%%%%%%%%%%%%%%%%%%%%%%%%%%%%%%%%%%%%%%%%%%%%%%%
Using values of $H_{C1}$(0) from Fig. \ref{Hc2} (a) and $\xi_{GL}(0)$, we obtained $\lambda_{GL}(0)$ values for Re$_7$B$_3$ and Re$_3$B, which are 191.88 $\pm$ 1.34~nm and 384.51 $\pm$ 1.53~nm, respectively.
We can determine the thermodynamic critical field $H_C(0) $ using the $\xi_{GL}(0)$  and $\lambda_{GL}(0)$ values using the equation,
\begin{equation}
    H_C(0) = \frac{\phi_0}{2\sqrt{2}\lambda_{GL}(0).\xi_{GL}(0)}.
\end{equation}
The $H_C(0)$ for Re$_7$B$_3$ is 0.184 $\pm$ 0.002~T while for  Re$_3$B  it is 0.171 $\pm$ 0.002~T. 

\begin{figure}[b]
\includegraphics[width=91.5mm]{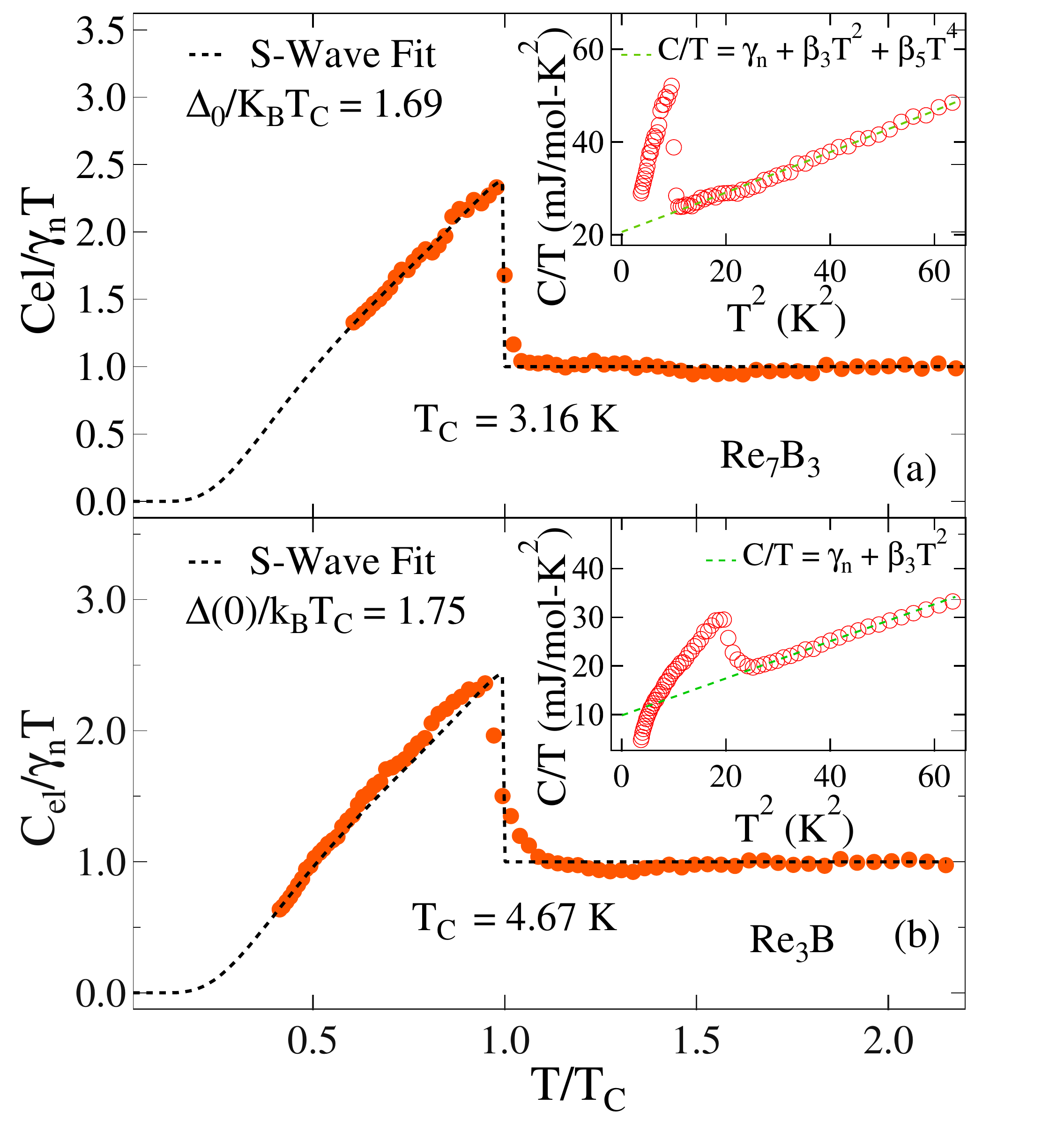}% Here is how to import EPS art
\caption{\label{SH}  The normalized electronic specific heat and its fit to the isotropic BCS model for (a) Re$_7$B$_3$ and (b) Re$_3$B. Insets:  C/T plotted against T$^2$ where the dotted green line represents the fits to the normal state specific heat.}
\end{figure}

The Ginzburg-Landau parameter $\kappa =\lambda_{GL}/\xi_{GL}$ can be calculated, giving 9.28 $\pm$ 0.12 for Re$_7$B$_3$ and 34.55 $\pm$ 0.38 for Re$_3$B. The $\kappa >1/\sqrt{2}$ as expected as both these compounds are type-II superconductors. 
%%%%%%%%%%%%%%%%%%%%%%%%%%%%%%%%%%%%%%%%%%%%%%%%%%%%%%%%%%%%%%%%%%%%%%%%%%%%%%%%%%%%%%%%%%%%%%%%
\paragraph*{Specific heat} Specific heat data for Re$_7$B$_3$ and Re$_3$B are shown in Fig. \ref{SH}. The dotted green line represents the fit to the specific heat data above T$_C$. The lattice contribution to the heat capacity above $T_C$ can  be written as $\beta_3T^3$ + $\beta_5T^5$ while the electronic portion is given by $\gamma_nT$. The specific heat (C) divided by temperature (T) was fit with C/T = $\gamma_n$ + $\beta_3T^2$ + $\beta_5T^4$ as shown in insets of Fig. \ref{SH} (a, b).   Here, $\gamma_n$ is the Sommerfeld coefficient and, $\beta_3$ and $\beta_5$ are the coefficients of the phononic contribution. From the fits of Re$_7$B$_3$, we obtained $\gamma_n$ = 20.61 $\pm$ 0.80~mJ/mol-K$^2$, $\beta_3$ = 0.42 $\pm$ 0.03~mJ/mol-K$^4$ and $\beta_5$ = 2.66 $\pm$ 0.19~$\mu$J/mol-K$^6$. Whereas, for Re$_3$B we obtained, $\gamma_n$ =  9.86 $\pm$ 0.27~mJ/mol-K$^2$, $\beta_3$ = 0.38 $\pm$ 0.01~mJ/mol-K$^4$ and $\beta_5$ = 0 (imposed).

 We can estimate the Debye temperatures ($\theta_D$) with the help of $\beta_3$ using \cite{PooleJr.2007},
\begin{equation}
\label{thetaD}
     \theta_D = \left(\frac{12\pi^4RN}{5\beta_3}\right)^\frac{1}{3},
\end{equation}
where R is the gas constant and N is the number of atoms per formula unit. We obtained $\theta_D$ = 359~K, 274~K for Re$_7$B$_3$ and Re$_3$B, respectively. Similarly, for non-interacting particles, the $\gamma_n$ relates to the density of states at Fermi level D$_C(E_f)$ through $\gamma_n = \left(\pi^2k_B^2/3\right)\times D_C(E_f)$. This yields D$_C(E_f)$ = 8.73 $\pm$ 0.34~$\frac{States}{eV f.u.}$ for Re$_7$B$_3$ and D$_C(E_f)$= 4.17 $\pm$ 0.11 $\frac{States}{eV f.u.}$ for Re$_3$B. 

 %%%%%%%%%%%%%%%%%%%%%%%%%%%%%%%%%%%%%%%%%%%%%%%%%%%%%%%%%%%%%%%%%%%%%
 %%%%%%%%%%%%%%%%%%%%%%%%%%%%%%%%%%%%%%%%%%%%%%%%%%%%%%%%%%%%%%%%%%%%%

We can determine the electron-phonon coupling constant ($\lambda_{e-ph}$) using the McMillan theory \cite{Mcmillan1967}. $\lambda_{e-ph}$ gives the strength of the attractive interaction between the electrons due to the interaction with the lattice. The following expression involves $\theta_D$, T$_C$ and the residual screened Coulomb interaction parameter ($\mu^*$).   
\begin{equation}
    \lambda_{e-ph} = \frac{1.04+\mu^*\ln(\theta_D/1.45T_C)}{(1-0.062\mu^*)\ln(1-0.062\mu^*)-1.04}
\end{equation}

where $\mu^*$ can be  assumed to be 0.13 as for many intermetallic superconductors \cite{Mcmillan1967}. The $\lambda_{e-ph}$ values obtained for Re$_7$B$_3$ (0.54) is close to other weak coupling non-centrosymmetric superconductors like TaOs \cite{Singh2017a}. For Re$_3$B, the value of $\lambda_{e-ph}$ (0.64) is comparable to other moderately coupled superconductors like Re$_6$Hf (0.63) \cite{Singh2016},  Re$_{24}$Ti$_5$ (0.6) \cite{ShangRe24Ti5, Lue2013}, and  Re$_3$Ta (0.62) \cite{TBarker2018} suggesting moderately strong coupling in Re$_3$B. 
 
The electronic portion of the heat capacity (C$_{el}$) can be obtained by subtracting the phononic contribution (C$_{ph}$) from the total heat capacity. The electronic heat capacity can be computed from the entropy using the thermodynamic relation \cite{Tinkham1996},
\begin{equation}
    \frac{S}{\gamma_nT_C} =\frac{-6}{\pi^2k_BT_C}\int_{0}^{\infty}\left[(1-f)\ln(1-f) + f\ln f\right]d\xi,
\end{equation} 
 where $f =  (1+e^{E/K_BT})^{-1}$ and $\xi$ is the energy of the normal electrons relative to the Fermi energy. The fermion excitation energy can be written as $E(\xi)=\left[\xi^2 +\Delta^2\right(t)]^{1/2}$ where energy gap, $\Delta$, has temperature dependence in accordance with the BCS approximation \cite{Carrington2003},
 
\begin{equation}
    \label{deltaT}
    \Delta(t) = \Delta_0\tanh(1.82\{1.018[(T_C/T)-1]\}^{0.51}),
\end{equation}

where $\Delta_0$ is the $\Delta$ at zero temperature. The normalized electronic heat capacity (C$_{el}/\gamma_nT$) is plotted against the reduced temperature (T/T$_C$) in Fig. \ref{SH} (a) and (c).
The dotted black line is the fit of the $\alpha$ model \cite{Johnston2013}, derived from BCS theory, for an isotropic gap superconductor as described above. The values of $\alpha=\Delta_0/k_BT_C$, the normalized gap,  as obtained from the fits are 1.69 for Re$_7$B$_3$ and  1.75 for Re$_3$B. These values are slightly smaller but very close to the BCS value $\alpha_{BCS}$= $\Delta_0/k_BT_C$ = 1.764.
The specific heat jump (C$_{el}/\gamma_nT_C$) at T$_C$, as determined from the specific heat plots are $\approx$ 1.348 for Re$_7$B$_3$ and  $\approx$ 1.428 for Re$_3$B. Both these values are also strikingly close to the BCS value (1.43) for conventional s-wave superconductors.

%%%%%%%%%%%%%%%%%%%%%%%%%%%%%%%%%%%%%%%%%%%%%%%%%%%%%%%%%%%%%%%%%%%%%%%%%%%%%%%%%%%%%%%%%%%%%%%%%
\paragraph*{Zero-field (ZF) muon spin relaxation/rotation ($\mu$SR)} 
We performed the ZF-$\mu$SR measurements to search for possible small spontaneous fields associated with the time-reversal symmetry breaking (TRSB) in the superconducting state. To investigate the change in internal fields between normal and superconducting state, we recorded the asymmetry spectra at several temperature points on both sides of the transition temperature and modeled it with the static Gaussian Kubo-Toyabe (KT)  Eq.  \ref{KT}. The Gaussian KT model \cite{Hayano1979} describes the relaxation due to a Gaussian distribution of random fields appropriate for densely packed random moments arising from nuclear dipoles. 
 \begin{figure}[t]
\includegraphics[width=92mm]{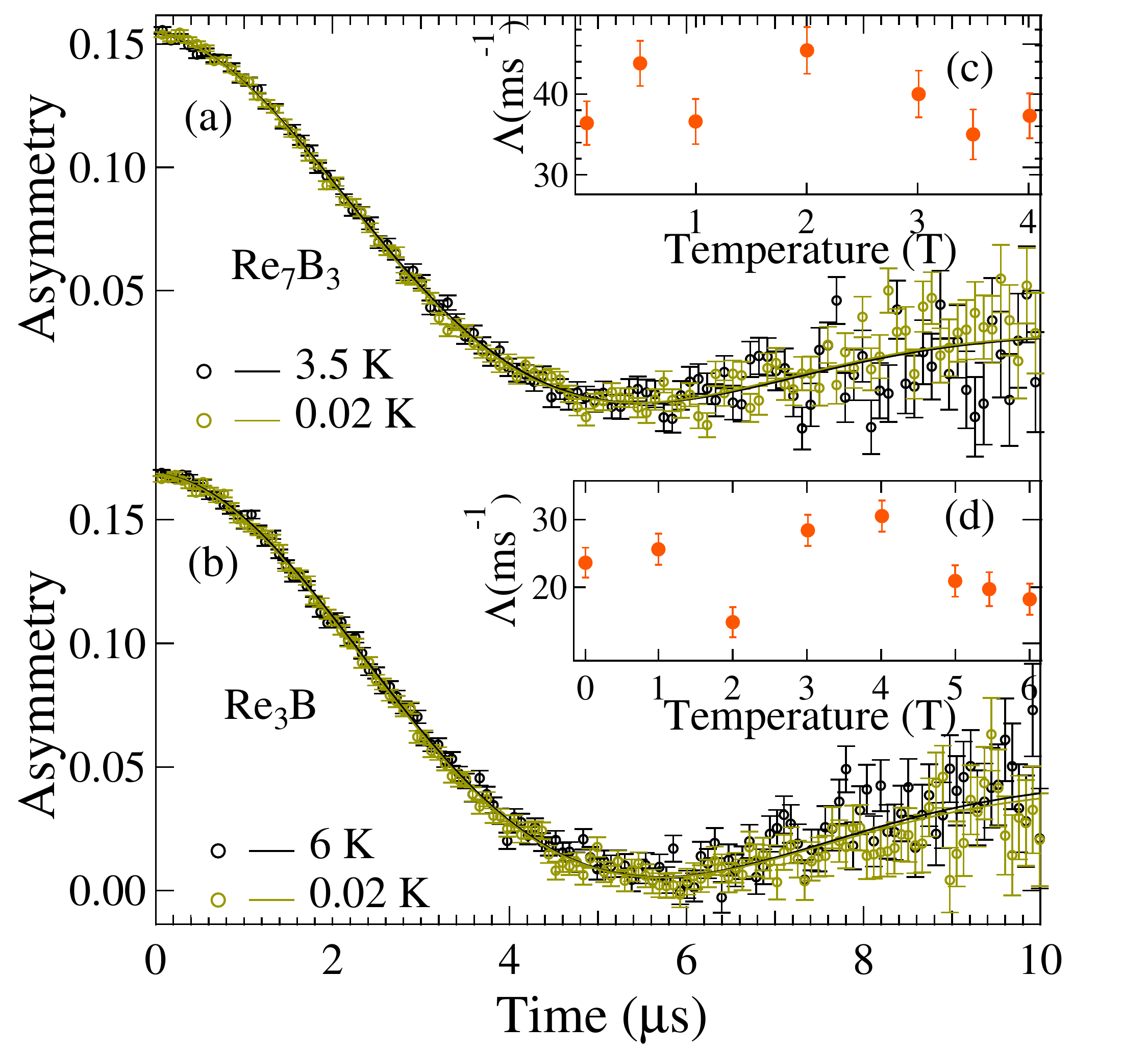}% Here is how to import EPS art
\caption{\label{ZF} (a) The representative zero-field spectra from Re$_7$B$_3$ in normal state (T= 3.5K) and superconducting state (T= 20~mK). (b) The ZF spectra obtained from Re$_3$B in normal state (T = 6~K) and superconducting state (T= 20~mK). (c) and (d) represent the electronic relaxation rate ($\Lambda$) obtained from the fits. }
\end{figure}

\begin{equation}
    \label{KT}
    G_{KT}(t) = \frac{1}{3} + \frac{2}{3}(1-\sigma_{ZF}^2)\exp\left(\frac{-\sigma_{ZF}^2t^2}{2}\right)
\end{equation}
where $\sigma_{ZF}$ is relaxation due to the static and randomly oriented nuclear dipole moments located in the vicinity of the muon site. $\sigma_{ZF}/\gamma_\mu$ is the width of the local field distribution and $\gamma_\mu$ is the gyromagnetic ratio. The complete asymmetry spectra was fit using the expression
\begin{equation}
    \label{KT1}
   A(t) = A_1G_{KT}(t)\exp\left(-\Lambda t\right) + A_{BG},
\end{equation}
where $A_1$ is the asymmetry of the sample signal. $A_{BG}$ is the asymmetry contributed by the background. The exponential relaxation ($e^{-\Lambda t} $) accounts for the relaxation due to possible spontaneous fields. This model was used to fit the asymmetry spectra collected at various temperatures on either side of T$_C$ and the resultant values of electronic relaxation rate ($\Lambda$) are depicted in the insets of Fig. \ref{ZF}. The fits of the asymmetry spectra trace almost the same function as we can see in Fig. \ref{ZF} (a-b). The Gaussian relaxation rate ($\sigma_{ZF}$) largely remained constant for different temperature scans and therefore, it was made a global variable during the fits. The values of $\sigma_{ZF}$ obtained for Re$_7$B$_3$ and Re$_3$B are 0.320 $\pm$ 0.001~$\mu s^{-1}$ and 0.304 $\pm$ 0.001~$\mu s^{-1}$, respectively, which correspond to nuclear dipole fields of $\approx$ 0.376~mT and $\approx$ 0.357~mT.   In addition, there is no visible trend in the $\Lambda$ down to the lowest temperature, rather, the $\Lambda$ values merely scatter around a temperature independent average value. The rms deviations of the $\Lambda$ corresponds to the fields of 0.0083~mT and 0.0075~mT for Re$_7$B$_3$ and Re$_3$B, respectively, which corresponds to an upper limit of any spontaneous magnetic field roughly 1/50$^{th}$ of the size of the nuclear dipole fields. These limits are smaller than most of the typical fields seen in TRS breaking superconductors between 0.01 and 0.05~mT \cite{Luke1993,Luke1998}. However, we do note that we cannot exclude TRS breaking fields less than about 0.01 mT which have been seen in few superconductors with smaller	nuclear dipole fields	 \cite{Singh2017, Zang2019}.

Our results are consistent with Shang et. al. \cite{ShangRe,Shang} inferences on correlation between average nuclear moment ($\mu_n$) and magnitude of the internal field, as well as the importance of critical rhenium concentration for TRS breaking in Re based superconductors. Re$_7$B$_3$ and Re$_3$B,  due to smaller nuclear magnetic moment of B and relatively low Re content, have lower average nuclear moments (2.84$\mu_N$ and 2.90$\mu_N$) than other TRS breaking superconductors, hence, show TRS breaking fields of value zero or less than 0.01 mT. This is irrespective of the presence or absence of inversion symmetry in their crystal structure. These results are in line with the recent work on iso-stoichiometric Re$_3$W \cite{Biswas2012} which exists in both centrosymmetric and non-centrosymmetric crystal structures and preserves time-reversal symmetry.

\begin{figure}[t]
\includegraphics[width=91mm]{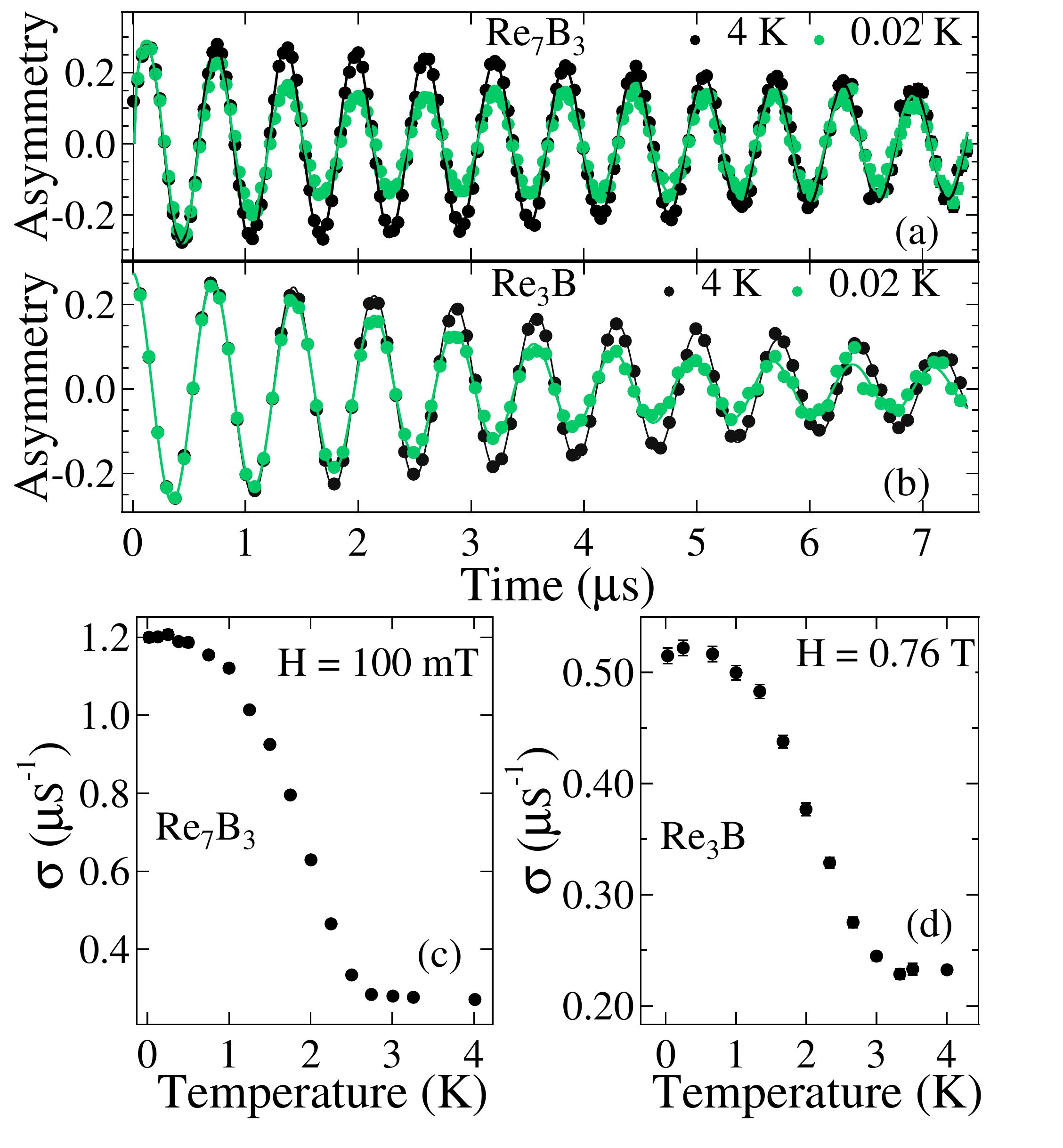}% Here is how to import EPS art
\caption{\label{TF} (a) The representative transverse-field (TF) $\mu$SR asymmetry spectra modified in a rotating reference frame with frequency (a) $\gamma_{\mu}$90~mT for Re$_7$B$_3$ (b) $\gamma_{\mu}$750~mT for Re$_3$B depicts the difference in relaxation rates ($\sigma$) between normal and superconducting state. The sample relaxation rate ($\sigma$) as obtained from the asymmetry data fits are plotted in  (c) for Re$_7$B$_3$ and  (d)  for Re$_3$B.
 } 
\end{figure}

%%%%%%%%%%%%%%%%%%%%%%%%%%%%%%%%%%%%%%%%%%%%%%%%%%%%%%%%%%%%%%%%%%%%%%%%%%%%%%%%%%%%%%%%%%%%%%%%%
\paragraph*{Transverse-field (TF) muon spin relaxation/rotation ($\mu$SR)} 
For our TF $\mu$SR experiment,  samples were cooled in fields 100~mT and 760~mT for Re$_7$B$_3$ and Re$_3$B, respectively, which is well above their lower critical fields ($H_{C1}$). The application of field results in a reduction of the T$_C$ in $\mu$SR, compared to low field magnetization and specific heat measurements. Asymmetry spectra were recorded above and below the superconducting transition temperatures. Rotating reference frame (rrf) plots of the asymmetry spectra are shown in Fig. \ref{TF} (a-b). The rrf field for Re$_7$B$_3$ and Re$_3$B was 90~mT and 750~mT, respectively.

The TF-$\mu$SR asymmetry signal decays with time due to the in-homogeneous field distribution due to the underlying flux line lattice (FLL) below T$_C$. Above T$_C$ the depolarization rate is lower but nonzero due to the Gaussian field distribution of  randomly oriented nuclear dipole moments. 
The time evolution of asymmetry can be described using two sinusoidally oscillating terms with Gaussian and exponential relaxation terms multiplied to it for sample and background (Ag Sample holder) part of the signal, respectively. 
\begin{equation}
\begin{split}
    G_{TF}(t) = &A[(1-F)e^{-\psi t}\cos(\omega_{bg} t + \phi) + \\ &(F) e^{\left(\frac{-\sigma^2 t^2}{2}\right)}\cos(\omega t + \phi)],
\end{split}
\end{equation}
 where $\omega$ and $\omega_{bg}$ are the frequencies of the muon precision signal in the sample  and background, respectively.  F is the fraction of the signal coming from the sample. A is the asymmetry and $\phi$ is the initial phase of muons entering the sample. $\sigma$ and $\psi$ are the sample and background depolarization rates, respectively. The $\psi$ value obtained from TF fits for Re$_7$B$_3$ is zero whereas for Re$_3$B it is 0.15 $\mu $s$^{-1}$ .  The temperature dependence of the depolarization/relaxation rate is obtained from the fits of the asymmetry spectra recorded at various temperatures as shown in Fig. \ref{TF}(c-d).
 
 In the case of superconductors, the muon depolarization rate ($\sigma$) consists of temperature independent background rate ($\sigma_{bg}$) and temperature dependent superconducting rate ($\sigma_{sc}$), $\sigma = \sqrt{\sigma_{bg}^2 + \sigma_{sc}^2}$. The superconducting depolarization rate ($\sigma_{sc}$) is the measure of the mean square in-homogeneity in the field, $\langle(\Delta B^2\rangle$, sampled by the muons due to the underlying flux line lattice \cite{Aeppli1987}, where, $\langle(\Delta B)^2\rangle = \langle(B-\langle B\rangle)^2\rangle$. The depolarization rate for vortex lattice is given by 
 \begin{equation}
 \label{variance}
     \sigma_{sc} = \langle(\Delta B)^2\rangle/\gamma_\mu  ,
 \end{equation}
 where $\gamma_\mu$ (= 2$\pi\times$135.5~MHz/T) is gyromagnetic ratio of the muon.
We can use $\sigma_{sc}$(T) to calculate the penetration depth ($\lambda$) as a function temperature. Brandt's approximation for the Abrikosov solution of the linearized GL theory gives Eq. \ref{Penedepth13}, which works well for the field range examined for Re$_3$B \cite{Brandt2003}.

\begin{equation}
    \label{Penedepth13}
    \frac{1}{\lambda^2(T)} = \frac{5.814\times\sigma_{sc}(T)}{\gamma_\mu\Phi_0\left(1-h\right)[1+1.21(1-\sqrt{h})^3]}
\end{equation}
where $\Phi_0$ is the magnetic flux quantum and h $\equiv$ $H/H_{C2}$, is temperature dependent reduced field.
For Re$_7$B$_3$ the ratio $H/H_{C2}$ is very small and therefore, $\lambda$ can be extracted using  the  simplified expression \cite{Brandt2003},
\begin{equation}
     \label{Penedepth11}
     \frac{1}{\lambda^2(T)} = \frac{\sigma_{sc}(T)}{0.061\times\gamma_\mu\Phi_0}.
\end{equation}
 
\begin{figure}[t]
\includegraphics[width=91mm]{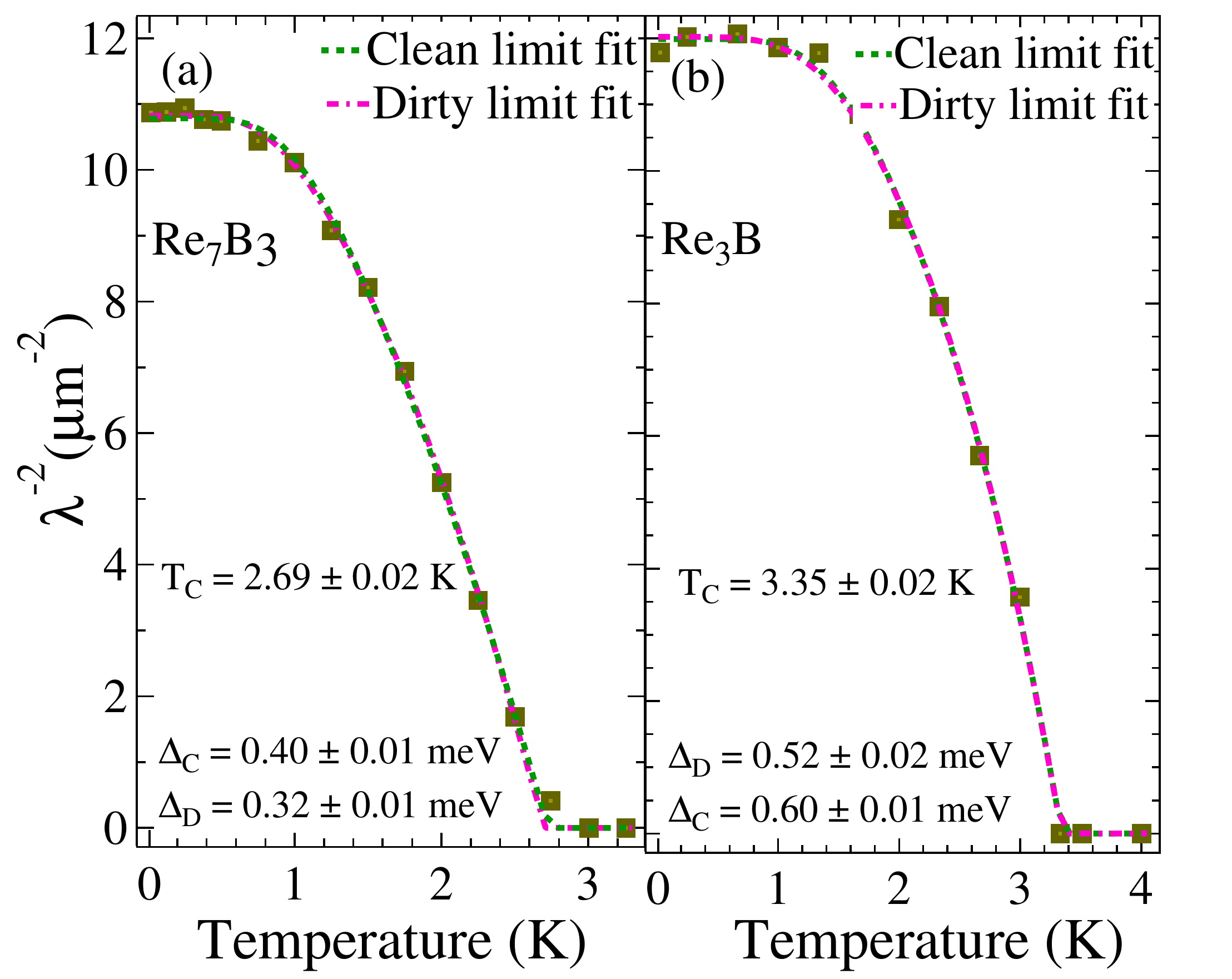}% Here is how to import EPS art
\caption{\label{TFfits} 
The graphs depicts the temperature dependence of inverse square penetration depth data calculated using (a) Eq. \ref{Penedepth11} for Re$_7$B$_3$ (b) Eq. \ref{Penedepth13} for Re$_3$B. The green and pink  dotted lines represent the clean and dirty limit BCS fits of the data. } 
\end{figure}
The London penetration depth, $\lambda$(0), for Re$_7$B$_3$ and Re$_3$B are 303.2 nm and 291.3 nm, which are comparable with the  Ginzburg-Landau penetration depth ($\lambda_{GL}(0)$) values obtained from the magnetization measurements. However, the values obtained $\mu$SR are more reliable, as the magnetic susceptibility was not corrected for demagnetization effects which mostly affects the extracted H$_{C1}$ values, where the magnetization is largest and therefore, the effects are more pronounced.

We can get more insight into the nature of the superconducting gap by modeling the London penetration depth obtained earlier, with the s-wave BCS superconductor model. This requires the knowledge of whether the superconductor is in the clean or dirty limit, which can be done by estimating the ratio of coherence length to the quasiparticle mean free path, $\xi$/$l$. If $\xi$/$l$ is a number much less than one, then the superconductor is considered to be in the clean limit. We estimated this ratio using the eq. 17-19 from Arushi et. al. \cite{Arushi2020}. We find that both Re$_7$B$_3$ and Re$_3$B are  clean limit superconductors with $\xi$/$l$ = 0.09 and 0.40, respectively.

In the  clean limit, the London penetration depth ($\lambda$) can be fit using the equation,
 \begin{equation}
     \label{clean}
     \frac{\lambda^{-2}(T)}{\lambda^{-2}(0)} = 1 + 2\int_{\Delta(T)}^{\infty}\frac{\partial f}{\partial E} \frac{EdE}{\sqrt{E^2-\Delta(T)}},
 \end{equation}
 where f = $\left[1+\exp(E/k_BT)\right]^{-1}$ and $\Delta(T)$ is given by  Eq.  \ref{deltaT}.
 For completeness, we can fit the results in the dirty limit as well, noting that we used normal state estimate for mean free path.   In the dirty local limit within the London approximation, we can write
\begin{equation}
    \label{DirtyPene}
    \frac{\lambda^{-2}(T)}{\lambda^{-2}(0)} = \frac{\Delta(T)}{\Delta(0)}\tanh\left[\frac{\Delta(T)}{2k_BT}\right],
\end{equation}
 where $\Delta(T)$ is given by Eq.  \ref{deltaT}, the BCS approximation for the energy gap. 
 The fits of the data in the two regimes are plotted in Fig. \ref{TFfits}. Both, clean and dirty limit fits are indistinguishable and fit the data well, implying the absence of nodes in superconducting gap for both.
 
 For Re$_7$B$_3$, the dirty limit BCS fit yields $\Delta_D(0)$ = 0.32 $\pm$ 0.01~meV while the clean limit fit yields $\Delta_C(0)$ = 0.40 $\pm$ 0.01~meV. Consequently, the normalized gap values are $\Delta_D/k_BT_C$ (1.36 $\pm$ 0.04) and $\Delta_C/k_BT_C$ (1.72 $\pm$ 0.05). Similarly, the dirty limit fit for Re$_3$B gives $\Delta_D(0)$ = 0.52 $\pm$ 0.02~meV while the clean limit fit yields $\Delta_C(0)$ = 0.60 $\pm$ 0.01~meV. The reduced gap values comes out to be $\Delta_D/k_BT_C$ (1.80 $\pm$ 0.08 ) and $\Delta_C/k_BT_C$ (2.07 $\pm$ 0.05).

%%%%%%%%%%%%%%%%%%%%%%%%%%%%%%%%%%%%%%%%%%%%%%%%%%%%%%%%%%%%%%%%%%%%%%%%%%%%%%%%%%%%%%%%%%%

\begin{table}[t]%The best place to locate the table environment is directly after its first reference in text
\caption{\label{tab:table1}%
Table presenting the superconducting and normal state parameter values for Re$_7$B$_3$ and Re$_3$B. 
}
\begin{ruledtabular}
\begin{tabular}{c c c c}
\textrm{Parameter}&
\textrm{unit}&
\multicolumn{1}{c}{\textrm{Re$_7$B$_3$}}&
\textrm{Re$_3$B}\\
\colrule
T$_C$ (Mag.) & K & 3.2 & 5.19\\
H$_{C1}(0)$ & mT & 9.98 $\pm$ 0.08 & 4.05 $\pm$ 0.03\\
H$_{C2}(0)$ & T & 0.78 $\pm$ 0.02 & 2.67 $\pm$ 0.04\\
H$_{C}(0)$ & T & 0.184 $\pm$ 0.002 & 0.171 $\pm$ 0.002\\
$\xi_{GL}(0)$ & nm & 20.68 $\pm$ 0.12 & 11.13 $\pm$ 0.08\\
$\Lambda_{e-ph}$ &  & 0.54 & 0.64\\
$\lambda_{GL}(0)$ (Mag.) & nm & 198.18  & 384.51 \\
$\lambda(0)$ ($\mu$SR) & nm & 303.2  & 291.3\\
$k_{GL}$ &  & 9.277 $\pm$ 0.12 & 34.55 $\pm$ 0.38\\
$\gamma_n$ & $mJ/mole-K^{2}$ & 20.61 $\pm$ 0.80 & 9.86 $\pm$0.27\\
$\theta_D$ & K & 359 & 274\\
$\xi/l$ &  & 0.09 & 0.40\\
$\Delta C_{el}/\gamma_n T_C$ &  & 1.348 & 1.428\\
$\Delta_0/k_BT_C$\footnote{Specific heat} &  & 1.69 & 1.75\\
$\Delta_0/k_BT_C$\footnote{TF $\mu$SR (clean limit)} & & 1.72 $\pm$ 0.05 & 2.07 $\pm$ 0.05\\ 
$\Delta_0/k_BT_C$\footnote{TF $\mu$SR (dirty limit)} & & 1.36 $\pm$ 0.04 & 1.80 $\pm$ 0.08\\ 

\end{tabular}
\end{ruledtabular}
\end{table}

%%%%%%%%%%%%%%%%%%%%%%%%%%%%%%%%%%%%%%%%%%%%%%%%%%%%%%%%%%%%%%%%%%%%%%%%%%%%%%%%%%%%%%%%%%%
\section{Conclusion}
We carried out detailed resistivity, magnetization, specific heat, TF-$\mu$SR, and ZF-$\mu$SR measurements on polycrystalline samples of Re$_7$B$_3$ and Re$_3$B. The comparison of their various superconducting and normal state parameters derived from the above measurements are listed down in Table~\ref{tab:table1}. Re$_7$B$_3$ is a weakly coupled superconductor while Re$_3$B is a moderately coupled superconductor. Both Re$_7$B$_3$ and Re$_3$B show conventional s-wave superconductivity with superconducting gaps of 0.40 $\pm$ 0.02~meV and 0.60 $\pm$ 0.02~meV, respectively in the clean limit. We find no evidence for TRS breaking; the magnitude of the maximum possible TRS breaking fields that might go undetected in Re$_7$B$_3$ (7.5~$\mu T$) and Re$_3$B (8.3~$\mu T$) are small in comparison to many TRS breaking superconductors (10~$\mu T$ - 50~$\mu T$) \cite{Luke1993, Luke1998}. Therefore, despite having many ingredients for unconventional superconductivity, Re$_3$B and Re$_7$B$_3$ seem to be quite conventional.

%%%%%%%%%%%%%%%%%%%%%%%%%%%%%%%%%%%%%%%%%%%%%%%%%%%%%%%%%%%%%%%%%%%%%%%%%%%%%%%%%%%%%%%%%%%
\section{Acknowledgments}
We would like to thank B.S. Hitti, S. Dunsiger, and G.D. Morris for their assistance during the $\mu$SR measurements. Work at McMaster was supported by the Natural Sciences and Engineering Research of Council of Canada. R. P. S. acknowledges the Science and Engineering Research Board, Government of India for the Core Research Grant CRG/2019/001028. Financial support from DSTFIST Project No. SR/FST/PSI-195/2014(C) is also thankfully acknowledged.
%%%%%%%%%%%%%%%%%%%%%%%%%%%%%%%%%%%%%%%%%%%%%%%%%%%%%%%%%%%%%%%%%%%%%%%%%%%%%%%%%%%%%%%%%%%

\nocite{*}

\bibliography{ReB}% Produces the bibliography via BibTeX.

\end{document}